\documentclass[sigconf, screen,nonacm]{acmart}
\usepackage{multirow}
\AtBeginDocument{%
  }

\setcopyright{none}
\begin{document}

\title[Multistakeholder Value-Driven Co-Design of Recommender System Evaluation Metrics in Digital Archives]{A Multistakeholder Approach to Value-Driven Co-Design of Recommender System Evaluation Metrics in Digital Archives}


\author{Florian Atzenhofer-Baumgartner}
\email{florian.atzenhofer-baumgartner@student.tugraz.at}
\orcid{0000-0001-8157-8629}
\affiliation{%
  \institution{Graz University of Technology}
  \city{Graz}
  \country{Austria}
}

\author{Georg Vogeler}
\email{georg.vogeler@uni-graz.at}
\orcid{0000-0002-1726-1712}
\affiliation{%
  \institution{University of Graz}
  \city{Graz}
  \country{Austria}
}

\author{Dominik Kowald}
\email{dkowald@know-center.at}
\orcid{0000-0003-3230-6234}
\affiliation{%
  \institution{Know Center Research GmbH}
  \city{Graz}
  \country{Austria}\\
  \institution{Graz University of Technology}
  \city{Graz}
  \country{Austria}
}


\begin{abstract}
This paper presents the first multistakeholder approach for translating diverse stakeholder values into an evaluation metric setup for Recommender Systems (RecSys) in digital archives. While commercial platforms mainly rely on engagement metrics, cultural heritage domains require frameworks that balance competing priorities among archivists, platform owners, researchers, and other stakeholders. To address this challenge, we conducted high-profile focus groups (5 groups × 5 persons) with upstream, provider, system, consumer, and downstream stakeholders, identifying value priorities across critical dimensions: visibility/representation, expertise adaptation, and transparency/trust. Our analysis shows that stakeholder concerns naturally align with four sequential research funnel stages: discovery, interaction, integration, and impact. The resulting evaluation setup addresses domain-specific challenges including collection representation imbalances, non-linear research patterns, and tensions between specialized expertise and broader accessibility. We propose directions for tailored metrics in each stage of this research journey, such as research path quality for discovery, contextual appropriateness for interaction, metadata-weighted relevance for integration, and cross-stakeholder value alignment for impact assessment. Our contributions extend beyond digital archives to the broader RecSys community, offering transferable evaluation approaches for domains where value emerges through sustained engagement rather than immediate consumption.
\end{abstract}




\maketitle

\begingroup
\renewcommand\thefootnote{}\footnotetext{%
  \hspace{-1.5em}\raisebox{5pt}{%
    \begin{minipage}[t]{\columnwidth}
      \footnotesize
      © Florian Atzenhofer-Baumgartner, Georg Vogeler, and Dominik Kowald, 2025. This is the author's version of the work entitled ``A Multistakeholder Approach to Value-Driven Co-Design of Recommender System Evaluation Metrics in Digital Archives''. It is posted here for your personal use, not for redistribution. The definitive version of record was accepted for publication in the \textit{19th ACM Conference on Recommender Systems (RecSys 2025)}, \url{10.1145/3705328.3748026}.
    \end{minipage}%
  }%
}
\endgroup

\section{Introduction}
Recommender Systems (RecSys) in cultural heritage contexts face unique challenges that conventional evaluation approaches—often developed for domains emphasizing immediate user engagement—
fail to fully address. While many platforms are moving beyond simple engagement metrics, digital archives, libraries, and repositories serve diverse stakeholders with divergent values that require specialized evaluation frameworks.

Digital archives like Monasterium.net—the largest host of medieval charters with over 685,000 documents—exemplify these challenges through their complex stakeholder ecosystem \cite{Atzenhofer-Baumgartner_Geiger_Vogeler_Kowald_2024}: upstream (archivists, curators, librarians); provider (aggregators, digitization services); system (owners, developers, moderators); consumer (researchers, educators, students); and downstream (publishers, educational platforms, media). These groups can hold fundamentally different perspectives on what constitutes ``good'' recommendations, creating an inherently multistakeholder evaluation problem.

While the RecSys community increasingly recognizes multistakeholder approaches \cite{Burke_Adomavicius_Bogers_Noia_Kowald_Neidhardt_Özgöbek_Pera_Tintarev_Ziegler_2025,dagstuhl_multi_2024}, translating stakeholder values into concrete evaluation metrics remains challenging, particularly in less-commercial domains. Our prior work has laid the foundation by identifying relevant stakeholders and values in the given domain \cite{Atzenhofer-Baumgartner_Geiger_Vogeler_Kowald_2024}, but the operational step of translating these insights into a structured evaluation setup remained an open challenge.

Our paper bridges this critical research gap through three main contributions:
\begin{enumerate}
    \item The \textit{first multistakeholder focus group study} in \textit{cultural heritage}, involving 25 high-profile domain experts selected from 70 candidates. Our approach answers recent calls for comprehensive stakeholder analysis by incorporating perspectives beyond RecSys end-users.
    
    \item The \textit{first value-driven RecSys evaluation setup} for \textit{digital archives} that addresses the challenges of less commercial recommendation contexts. Our metrics respond to specific needs of cultural heritage domains where value emerges through extended scholarly engagement.
    
    \item A systematic approach for \textit{generating RecSys evaluation metric setups} that identifies key measurement dimensions and value trade-offs in domains with complex (scholarly) information needs, while providing flexible implementation paths adaptable across digital archives, libraries, repositories.
\end{enumerate}

These contributions can significantly advance RecSys for cultural heritage institutions in general. By prioritizing stakeholder values over simplistic engagement metrics, our work provides a foundation for systems that balance competing priorities in domains where recommendation value emerges through extended stakeholder engagement rather than immediate consumption.

\section{Related Work and Background}
There are three lines of research related to our work: 

\vspace{1mm} \noindent \textbf{Multistakeholder RecSys.} 
Recent research has highlighted limitations of single-objective RecSys, with multistakeholder approaches gaining traction across various domains. In news recommendation, Vandenbroucke and Smets \cite{Vandenbroucke_Smets_2024} identify disconnects between commercial metrics and stakeholder values, while Van Den Bogaert et al. \cite{Van_Den_Bogaert_Geerts_Harambam_2024} explore co-designing scaffolds for higher user agency. Similar challenges have been identified in tourism recommendation systems \cite{Banerjee_2023}, where balancing the needs of tourists, local communities, and environmental sustainability requires careful consideration of several aspects, including societal impacts.

Burke et al. \cite{Burke_Adomavicius_Bogers_Noia_Kowald_Neidhardt_Özgöbek_Pera_Tintarev_Ziegler_2025} provide a comprehensive multistakeholder framework considering needs of consumers, providers, platform owners, and society, while Smith et al. \cite{Smith_Satwani_Burke_Fiesler_2024} emphasize that provider fairness requires different metrics than consumer fairness. While domains like job matching or educational platforms also deal with long-term goals, cultural heritage archives present a unique challenge where value emerges from an extended and iterative (scholarly) workflow of interpretation, synthesis, and re-contextualization. This process-oriented value creation distinguishes it from many other domains and necessitates evaluation approaches that move beyond immediate interaction metrics. However, despite this growing body of work, cultural heritage remains largely unexplored from a multistakeholder perspective, with our recent work beginning to identify stakeholder tensions in digital archives, highlighting conflicts, e.g., between scholarly utility, collection representation, and platform growth \cite{Atzenhofer-Baumgartner_Geiger_Vogeler_Kowald_2024}.

\vspace{1mm} \noindent \textbf{From Values to Metrics.} 
Translating stakeholder values into evaluation metrics presents a persistent challenge. Stray et al. \cite{Stray_Halevy_Assar_Hadfield-Menell_Boutilier_Ashar_Bakalar_Beattie_Ekstrand_Leibowicz_et_al._2024} catalogs 36 human values in algorithm design, noting many lack established measurement approaches and require domain-specific operationalization. In RecSys, Deldjoo et al. \cite{Deldjoo_Jannach_Bellogin_Difonzo_Zanzonelli_2024} has reviewed fairness-aware recommendation, highlighting tensions between different fairness concepts and challenges in selecting metrics that meaningfully represent normative goals. De Biasio et al. \cite{De_Biasio_Montagna_Aiolli_Navarin_2023} has emphasized the need for metrics beyond accuracy to model recommendation impacts on human values. 

Comprehensive evaluation requires multiple dimensions simultaneously. Zangerle and Bauer \cite{Zangerle_Bauer_2023} introduce a Framework for Evaluating RecSys (FEVR), organizing the evaluation design space and offering guidance on appropriate evaluation configurations. Despite ample algorithmic approaches, Bauer et al. \cite{Bauer_Bagchi_Hundogan_Van_Es_2024} find that conceptual frameworks for measuring values remain underdeveloped in news recommendation systems. This gap is arguably significant in cultural heritage domains, where conventional engagement metrics fail to capture long-term utility and impact.

\vspace{1mm} \noindent \textbf{Digital Libraries and Scholarly Recommendation.}  
Digital libraries represent an important ``precursor'' to digital archives, with a lot of research on user needs and evaluation approaches. While having different foci, both face similar challenges in resource discovery. Dobreva et al. \cite{Dobreva_ODwyer_Feliciati_2012} highlight the importance of user-centered approaches, while Marchionini et al. \cite{Marchionini_Plaisant_Komlodi_2003} has long established frameworks addressing tensions between expert and novice users. 

A focus on considering diverse stakeholder needs has been recently alluded to in digital libraries contexts \cite{Marchionini_2024}, simultaneously addressing digital archives. Our work addresses this gap through a comprehensive stakeholder analysis in the form of focus groups. Connected to this, scholarly RecSys face unique challenges that commercial systems do not. Zhang et al. \cite{Zhang_Patra_Yaseen_Zhu_Sabharwal_Roberts_Cao_Wu_2023} find content-based filtering dominates this domain due to specific information needs, while Champiri et al. \cite{Champiri_Shahamiri_Salim_2015} emphasize how situational factors critically affect research-oriented recommendations. Biases in these systems have received increasing attention, with F\"arber et al. \cite{Färber_Coutinho_Yuan_2023} distinguishing between human-originated and system-induced biases that inform our collection representation evaluation metrics. 

Recent work by Wecker et al. \cite{Wecker_Kuflik_Goldberg_Lanir_Tabashi_2023} suggests a broader perspective on RecSys in cultural heritage, arguing for systems that go beyond individual user interests to consider societal impact and encourage exploration of diverse perspectives. This aligns with our multistakeholder approach, emphasizing the need to balance individual research needs with broader cultural heritage goals.

Our work extends our previous identification of challenges in RecSys for historical research \cite{Atzenhofer-Baumgartner_Geiger_Trattner_Vogeler_Kowald_2024} by developing a conceptual structure that translates the identified stakeholder values into concrete RecSys evaluation setups and towards metrics.

\section{Methodology}
Our study is grounded in a participatory, value-driven approach, aligning with principles from value-sensitive design where stakeholder values are a primary driver in the design and evaluation process. In the following, we describe our focus group and analysis procedures:

\vspace{1mm} \noindent \textbf{Participatory Focus Groups.} 
We conducted structured focus groups with five stakeholder groups (upstream, provider, system, consumer, and downstream) adapted from prior RecSys research to the digital archives domain. From a pool of over 70 candidates, we selected 25 high-profile experts who were representative of the domain and had connections to Monasterium.net. 

The selection process, guided by a semi-automated scoring approach, aimed to ensure role diversity and gender balance across groups. Before participation, all participants provided written informed consent, following institutional and ACM ethical guidelines. Each of the five groups included five experts, partly representing major organizations affiliated with Monasterium.net, ensuring findings generalize to broader contexts.
Table~\ref{tab:stakeholders} gives an overview of our 25 participants, and to foster transparency and reproducibility \cite{semmelrock2025reproducibility}, we share detailed (anonymized) information, including consent forms and discussion guides, in our repository\footnote{\url{https://github.com/atzenhofer/multistakeholder-archives-recsys}}.

\begin{table}[h]
\caption{An overview of our stakeholder groups, including the roles of the 25 participants of our study. While provider, system, and consumer stakeholder directly interact with the platform and its recommendations, upstream and downstream stakeholder almost never directly interact with them, but may be impacted by the outcomes.}
\label{tab:stakeholders}
\begin{tabular}{ll}
\hline
\textbf{Group} & \textbf{Roles} \\
\hline
Upstream (U1-U5) & Archivist, Curator, Librarian \\
Provider (P1-P5) & Manager, Specialist, Researcher \\
System (S1-S5) & Developer, Director, Specialist \\
Consumer (C1-C5) & Researcher, Educator, Student \\
Downstream (D1-D5) & Publisher, Editor, Technologist \\
\hline
\end{tabular}
\end{table}

\vspace{1mm} \noindent \textbf{Sessions} (March 2025) with scenario-based discussions lasted 60 minutes, were recorded, and transcribed using aTrain \cite{Haberl_Fleiß_Kowald_Thalmann_2024}/Whisper \cite{Radford_Kim_Xu_Brockman_McLeavey_Sutskever_2022}. Discussion questions centered on three topics from previously identified priority areas \cite{Atzenhofer-Baumgartner_Geiger_Vogeler_Kowald_2024}: visibility/representation, expertise adaptation, and transparency/trust. 

We included provocative statements (e.g., ``I don't need to understand why something is recommended as long as it's relevant to my research'') to elicit varied value positions through relevant scenarios \cite{Starke_Vrijenhoek_Michiels_Kruse_Tintarev_2024}.

\vspace{1mm} \noindent \textbf{Analysis and Development of the Evaluation Setup.} 
We analyzed transcripts using an abductive coding approach \cite{Vandenbroucke_Smets_2024}, combining deductive coding (using pre-established value categories from literature with inductive analysis of emerging patterns. The evaluation spaces and metric setups emerged directly from stakeholder discussions; our contribution was to formalize these insights. Initial analysis revealed recurring references to research stages, prompting a secondary coding round examining how values related to these stages. While participants described their workflows in phases, our synthesis structured these narratives into the four-part ``research funnel'' (discovery, interaction, integration, and impact). This emergent model provides a structure for organizing evaluation setups grounded in stakeholders' concerns.

\section{Results}

\begin{table*}[t]
  \caption{An overview of the proposed evaluation setup, mapping research funnel stages to key metric directions. This funnel reflects the scholarly workflow in digital archives as a progressive value creation process through four sequential stages: \textit{Discovery} (initial exploration and serendipitous finding), \textit{Interaction} (adaptive engagement and user control), \textit{Integration} (document synthesis and relationship analysis), and \textit{Impact} (long-term research outcomes and stakeholder value alignment). Each stage addresses distinct evaluation challenges while all stakeholders maintain interests across stages; metric directions are derived from stakeholder discussions to systematically structure what and how to measure recommendation value.}

  \label{tab:evaluation_setup}
  \small
  \begin{tabular*}{\textwidth}{@{\extracolsep{\fill}} l l p{11.5cm}}
  \toprule
  \textbf{Stage} & \textbf{Metric Direction} & \textbf{Description \& Addressed Values} \\
  \midrule
  \multirow{2}{*}{Discovery} & Research Path Quality & Evaluates recommendation sequences for coherence, balancing research focus with discovery. \\
   & Collection Representation & Measures recommendation distribution to counter biases from historical digitization priorities. \\
  \midrule
  \multirow{2}{*}{Interaction} & Contextual Appropriateness & Assesses adaptation to users' evolving (research) contexts and expertise to ensure relevance. \\
   & Control Effectiveness & Evaluates user controls' impact on quality and system learning, promoting user agency and trust. \\
  \midrule
  \multirow{2}{*}{Integration} & Metadata-Weighted Relevance & Adapts relevance metrics to metadata quality, ensuring accuracy by mitigating data-induced bias. \\
   & Document Relationship Insight & Measures support for discovering connections between documents to aid research synthesis. \\
  \midrule
  \multirow{2}{*}{Impact} & Research Integration & Assesses long-term value by tracking the inclusion of recommended items in research outputs. \\
   & Cross-Stakeholder Value Alignment & Measures balanced performance across stakeholder priorities through multi-objective evaluation. \\
  \bottomrule
  \end{tabular*}
  \end{table*}
\subsection{Key Value Priorities and Controversies}
We identify value patterns across functional, experience, and responsibility dimensions, as well as fundamental controversies:

\noindent \textbf{Universal Values.} All stakeholders emphasize \textit{relevance} and \textit{context awareness}, though with varying interpretations. \textit{Transparency} and \textit{trust} receive broad support, particularly from consumer and downstream groups concerned with scholarly credibility.

\vspace{1mm} \noindent \textbf{Collection Representation vs. Quality Highlighting.} Controversy emerges regarding whether to prioritize equal representation of collections or visibility for high-quality items. An upstream archivist (U2) states, ``I never understood Monasterium.net as a kind of competition between archives or collections, so I always thought about it as a tool for opening up archives'', reflecting a different focus than perspectives emphasizing notable documents.

\vspace{1mm} \noindent \textbf{Popularity vs. Scholarly Merit.} Popularity-based recommendations face strong rejection across groups. A downstream stakeholder (D3) states: ``Why do I care if the charter is trending or not?'', and another (D2) notes, ``If you have something that nobody looked at the last 150 years, I think this is at least one criteria for interesting research questions''. This reflects a fundamental difference between scholarly and commercial recommendation approaches.

\vspace{1mm} \noindent \textbf{Scholarly Accuracy vs. Broader Accessibility.} Tension emerges between maintaining scholarly integrity and accessibility. Upstream stakeholders debate whether simplifying interfaces sacrifices scholarly value, with one (U3) arguing, ``I don't really think you're sacrificing accuracy by having a simpler initial interface'', while another (U4) emphasizes, ``The crucial question would be: What is the aim of Monasterium.net? As I understand it and use it until now, it is for scholarly purposes''.

\vspace{1mm} \noindent \textbf{System Learning vs. User Control.} Different views emerge on balancing adaptive behavior and explicit control. A consumer stakeholder (C1) notes, ``If I get the impression that the system is learning with my input, I would make the efforts to put in my opinion'', emphasizing implicit personalization benefits. In contrast, system stakeholders prefer explicit control mechanisms, advocating for configurable interfaces with multiple recommendation modes; this is a clear tension regarding RecSys and interface design.

\vspace{1mm} \noindent \textbf{Research Focus vs. Serendipity.} Tension between focused research and serendipitous discovery is evident among consumer stakeholders. While one researcher (C2) states, ``I'm just looking for specific charters, and I don't want to see very interesting things without any of those features,'' another (C3) emphasizes, ``I think we need constantly to be challenged, so that we don't pursue the wrong, where we think we're going.'' An upstream stakeholder (U5) highlights unexpected discoveries: ``The most beautiful finds you make are the findings that you would not expect, and you will [rather] find them by equal representation''.

\vspace{1mm} \noindent \textbf{Metadata Quality and Bias.} Providers and system stakeholders demonstrate awareness of metadata quality variations and structural biases. A consumer (C5) notes, ``The main problem I see here is the available data at Monasterium.net, because there is, of course, a structural bias,'' while a provider (P1) observes, ``It moves the focus away from very interesting areas, it basically just perpetuates the interests of the scientists of the 19th and early 20th century.''

\vspace{1mm} \noindent \textbf{Trust through Transparency.} Trust emerges critical to explanation and transparency. An upstream archivist (U2) states, ``Trust and transparency is a key value for us. And that's why I always want to understand why something is recommended for me,'' while a system stakeholder (S3) notes, ``In my experience with academics, they frequently ask: ``Why am I getting this recommendation? I need to understand how the system works''.

\subsection{Research Funnel Alignment}
Stakeholder concerns aligned with four sequential stages of the scholarly research funnel, coinciding with information seeking behavior and progressive stages of researcher involvement \cite{Li_Zhang_Wang_2024}:

\vspace{1mm} \noindent \textbf{Discovery Stage.} Upstream stakeholders emphasized serendipity and unexpected connections, while providers showed awareness of structural biases. The system perspective was captured by S3: ``You can think about different stages of interaction - new users not too familiar with what's inside might benefit most from popularity, and see what is most known about that.'' demonstrating how recommendation approaches should adapt to different expertise levels or research stages.

\vspace{1mm} \noindent \textbf{Interaction Stage.} System stakeholders offered sophisticated understanding of progressive disclosure and adaptive interfaces. S1 emphasized content relationships: ``I'm currently missing one specific part, and that is the similarity of documents [...] based on content.'' Consumers emphasized control and agency, with C1 stating, ``For me, the central topic is relevance. Whatever I get, it must interest me. If not, I lose interest in the recommendations.''

\vspace{1mm} \noindent \textbf{Integration Stage.} Consumer stakeholders focused on document relationship understanding and research context maintenance. A system stakeholder (S4) hypothesized long-term goals: ``If we think deeper about the user: they're not interested in items; they explore a specific question and they look for historical evidence.'' Downstream stakeholders highlighted challenges in understanding contexts, with D5 noting, ``We don't really know which topic people are studying - are they using sources for historical analysis, for linguistics, for material questions?''

\vspace{1mm} \noindent \textbf{Impact Stage.} System stakeholders raised questions about attribution and collaboration, with S4 asking, ``There is also the question: if researchers have done some exploration of sources, and discovered something, should they count Monasterium.net and the RecSys as a collaborator?'' Provider stakeholders emphasized metrics beyond engagement, noting the challenge of popularity metrics in specialized collections that at several levels suffer from data sparsity.

\subsection{Proposed Evaluation Setup}
Based on our cross-stakeholder analysis, we propose an evaluation setup (summarized in Table \ref{tab:evaluation_setup}) organized around the four stages of a scholarly research funnel. Unlike conventional RecSys metrics that focus on immediate consumer satisfaction, our setup addresses cultural heritage domains where recommendation value unfolds across extended research journeys. For each stage, we identify metric directions that reflect shared stakeholder values or provide mechanisms to balance competing priorities—an advancement for domains where traditional engagement metrics fall short.

For the discovery stage, we propose \textit{Research Path Quality}, which evaluates research trajectories rather than isolated item relevance, measuring topic coherence and informational progression through sequences of recommendations. This relies on and extends session-based metrics \cite{Wang_Zhang_Hu_Zhang_Wang_Aggarwal_2022} and sequential recommendation approaches \cite{Wang_Zhang_Hu_Zhang_Wang_Aggarwal_2022} to scholarly contexts where research develops across multiple interactions. \textit{Collection Representation} addresses structural biases by measuring distribution of recommendations across contributing archives, time periods, and document types. This requires identifying imbalances relative to collection composition, addressing the structural biases that arise from historical digitization priorities, which are distinct from, though related to, algorithmic popularity bias. It connects to work on item coverage \cite{Zangerle_Bauer_2023} and fairness, including popularity bias mitigation \cite{Klimashevskaia_Jannach_Elahi_Trattner_2024,forster2025exploring}, but extends these to incorporate given humanities-specific requirements. In the context of digital humanities, researcher coverage is a key metric for assessing how the ``cold-start problem'' impacts the system's ability to recommend pertinent scholarly resources to researchers new to the platform \cite{lacic2015tackling}. 

For the interaction stage, we propose \textit{Contextual Appropriateness}, which assesses how recommendations align with shifting research contexts, measuring adaptation to evolving queries and expertise development in real-time. This extends beyond simple relevance to evaluate recommendations' responsiveness to user behavior. We also propose \textit{Control Effectiveness}, which evaluates how user control mechanisms (filtering, refinement, rejection) impact recommendation quality over time, measuring both immediate adaptation and system learning across sessions and contexts. These metrics align with intent-aware and context-aware RS \cite{Jannach_Zanker_2024}, emphasizing adaptation to shifting scholarly contexts over pure recommendation accuracy.

For the integration stage, we propose to use \textit{Metadata Quality Weighted Relevance}, which adapts traditional precision/recall metrics to account for varying metadata completeness across collections, addressing concerns about algorithmic bias toward better-documented materials. We also propose \textit{Document Relationship Insight}, which measures how effectively recommendations support comparative analysis between documents, extending beyond item-level relevance to evaluate support for scholarly synthesis and argument development. These metrics connect to graph-based and knowledge-enhanced recommendation approaches \cite{Gao_Zheng_Li_Li_Qin_Piao_Quan_Chang_Jin_He_et_al._2023}, addressing the gap in traditional evaluation approaches that assume uniform data quality—an assumption that fails in cultural heritage contexts due to significant variations in digitization quality.

For the impact stage, we propose \textit{Research Integration}, which measures how recommended items are incorporated into scholarly outputs, including citations, teaching materials, curated collections, or acknowledgments. This extends evaluation beyond immediate interactions to assess the system's longer-term influence on knowledge creation. We also propose \textit{Multistakeholder Value Alignment}, which measures balanced optimization across competing priorities using weighted satisfaction metrics based on stakeholder values. These metrics connect to multi-objective optimization approaches in RecSys \cite{jannach2023survey}, making trade-offs between priorities explicit and measurable while recognizing that in scholarly contexts, value often emerges only through extended engagement.

\subsection{Implementation Paths and Discussion}
Our approach addresses limitations in current RecSys evaluation while building upon established research directions. We discuss implementation paths and implications for cultural heritage domains and RecSys generally.

\vspace{1mm} \noindent \textbf{User Agency and Explainability.} 
A central theme was the importance of user agency and transparency. Focus group participants emphasized that researchers need to understand recommendations and maintain control. This aligns with findings that explanation and control affect both experience and recommendation quality \cite{Rezk_Simkute_Luger_Vines_Elsden_Evans_Jones_2024, Wardatzky_Inel_Rossetto_Bernstein_2025}. 
Intent-aware systems enhance user agency through multiple interest profiles \cite{Lian_Batal_Liu_Soni_Kang_Wang_Xie_2021}, intent signaling, adjustable parameters, and context-appropriate explanations. Control Effectiveness metrics address how these mechanisms support scholarly research; for instance, by measuring how often users engage with refinement tools and whether this leads to higher long-term satisfaction. We acknowledge that measuring control effectiveness is inherently challenging, yet it was deemed critical by stakeholders for fostering trust and supporting scholarly rigor.

\vspace{1mm} \noindent \textbf{Temporal and Context-Aware Evaluation.} 
Scholarly research involves extended timelines and shifting contexts that traditional evaluation approaches miss. The research funnel provides a structure to address these temporal dimensions by evaluating support for phase transitions, adaptation to evolving questions, and contribution to longer-term goals. 

This addresses a key limitation of current RecSys evaluation focused on immediate consumption. Our approach acknowledges that recommendation value often emerges through extended engagement and workflow integration. Metrics like Research Integration are therefore proposed to capture this long-term impact, even though their implementation requires complex, longitudinal data collection beyond typical RecSys evaluation.

\vspace{1mm} \noindent \textbf{Balancing Stakeholder Priorities.} 
Multistakeholder recommendation involves trade-offs between competing objectives. Our approach makes these trade-offs explicit through multi-objective optimization, configurable dashboards, and value alignment tracking. This approach could consider item, collection, process, and impact-level performance simultaneously, which addresses the need for deeper value integration in RecSys \cite{Stray_Halevy_Assar_Hadfield-Menell_Boutilier_Ashar_Bakalar_Beattie_Ekstrand_Leibowicz_et_al._2024}.

Our approach extends beyond cultural heritage to other domains characterized by complex stakeholder ecosystems, long-term value creation, and quality heterogeneity, such as scholarly publishing, job recommendation, and news media, where trustworthiness and sustained engagement arguably matter more than immediate clicks. What makes cultural heritage distinct is the intensive reinterpretation and integration of items into extended scholarly workflows—users do not just consume items but revisit, analyze, and synthesize them.

This creates unique evaluation challenges where value emerges through process, a core issue our research funnel addresses. This work thus offers transferable methods for mapping stakeholder values to concrete evaluation setups in other domains that acknowledge the importance of process and long-term value creation.

\section{Conclusion and Future Work}
This paper presents a multistakeholder approach for translating values into evaluation metrics for RecSys in cultural heritage contexts. By synthesizing insights from diverse stakeholder groups—including archivists, platform owners, researchers, and publishers—we connect abstract value discussions with concrete metric implementation. Our work organizes stakeholder values along a research funnel, providing a foundational structure for developing metrics that address the specific challenges of cultural heritage domains.

Our contributions extend beyond abstract discussions to practical evaluation approaches. The metric setups of research path quality, contextual appropriateness, and value alignment reimagine traditional RecSys evaluation to address scholarly processes, collection biases, and professional ethics. This setup balances competing priorities while prioritizing scholarly utility: it can be used as a blueprint for recommendation systems that respect both information-seeking needs and institutional missions. 

While developed for digital archives, our approach has broader implications for domains where recommendations support complex decision-making processes. By connecting stakeholder values to concrete metrics, we successfully demonstrate how participatory approaches can transform abstract values into practical evaluation setups—an issue the RecSys community continues to address. Ultimately, this work serves as a call for a more contextualized and value-conscious approach to RecSys evaluation, one that measures success not just by clicks, but by meaningful, long-term impact.

\vspace{1mm} \noindent \textbf{Limitations and Future Work.} 
Some limitations require further investigation. First, our stakeholder sample focused on domain experts but underrepresents non-expert users and third-party stakeholders. Second, the multi-role nature of participants creates challenges for stakeholder-specific value analysis that future studies should address. Third, our proposed metric directions require validation through implementation and user studies to assess their correlation with actual stakeholder satisfaction.

Future work should implement these metrics in real-world cultural heritage institutions, explore additional stakeholder perspectives, and adapt this approach to other contexts beyond scholarly research. The proposed structure could benefit from recent work on reproducibility challenges \cite{Shehzad_Dacrema_Jannach_2025} and joint modeling approaches that integrate multiple aspects of user behavior and content \cite{Zhao_Wang_Chen_Gao_Wang_Li_Jia_Liu_Guo_Tang_2025}.

\begin{acks}
This research is supported by the ERC Advanced Grant project (101019327) ``From Digital to Distant Diplomatics'' and the Austrian FFG COMET program. Special thanks to the Monasterium.net and ICARus team, and everyone across the cultural heritage community who have provided invaluable feedback and support throughout this research.
\end{acks}

\balance
\bibliographystyle{ACM-Reference-Format}
\bibliography{references}


\begin{thebibliography}{37}


\ifx \showCODEN    \undefined \def \showCODEN     #1{\unskip}     \fi
\ifx \showISBNx    \undefined \def \showISBNx     #1{\unskip}     \fi
\ifx \showISBNxiii \undefined \def \showISBNxiii  #1{\unskip}     \fi
\ifx \showISSN     \undefined \def \showISSN      #1{\unskip}     \fi
\ifx \showLCCN     \undefined \def \showLCCN      #1{\unskip}     \fi
\ifx \shownote     \undefined \def \shownote      #1{#1}          \fi
\ifx \showarticletitle \undefined \def \showarticletitle #1{#1}   \fi
\ifx \showURL      \undefined \def \showURL       {\relax}        \fi
\providecommand\bibfield[2]{#2}
\providecommand\bibinfo[2]{#2}
\providecommand\natexlab[1]{#1}
\providecommand\showeprint[2][]{arXiv:#2}

\bibitem[Atzenhofer-Baumgartner et~al\mbox{.}(2024a)]%
        {Atzenhofer-Baumgartner_Geiger_Trattner_Vogeler_Kowald_2024}
\bibfield{author}{\bibinfo{person}{Florian Atzenhofer-Baumgartner}, \bibinfo{person}{Bernhard~C. Geiger}, \bibinfo{person}{Christoph Trattner}, \bibinfo{person}{Georg Vogeler}, {and} \bibinfo{person}{Dominik Kowald}.} \bibinfo{year}{2024}\natexlab{a}.
\newblock \showarticletitle{Challenges in Implementing a Recommender System for Historical Research in the Humanities}.
\newblock  \bibinfo{number}{arXiv:2410.20909} (\bibinfo{date}{Oct.} \bibinfo{year}{2024}).
\newblock
\href{https://doi.org/10.48550/arXiv.2410.20909}{doi:\nolinkurl{10.48550/arXiv.2410.20909}}
\newblock
\shownote{arXiv:2410.20909}.


\bibitem[Atzenhofer-Baumgartner et~al\mbox{.}(2024b)]%
        {Atzenhofer-Baumgartner_Geiger_Vogeler_Kowald_2024}
\bibfield{author}{\bibinfo{person}{Florian Atzenhofer-Baumgartner}, \bibinfo{person}{Bernhard~C. Geiger}, \bibinfo{person}{Georg Vogeler}, {and} \bibinfo{person}{Dominik Kowald}.} \bibinfo{year}{2024}\natexlab{b}.
\newblock \showarticletitle{Value Identification in Multistakeholder Recommender Systems for Humanities and Historical Research: The Case of the Digital Archive Monasterium.net}.
\newblock  \bibinfo{number}{arXiv:2409.17769} (\bibinfo{date}{Sept.} \bibinfo{year}{2024}).
\newblock
\href{https://doi.org/10.48550/arXiv.2409.17769}{doi:\nolinkurl{10.48550/arXiv.2409.17769}}
\newblock
\shownote{arXiv:2409.17769}.


\bibitem[Banerjee(2023)]%
        {Banerjee_2023}
\bibfield{author}{\bibinfo{person}{Ashmi Banerjee}.} \bibinfo{year}{2023}\natexlab{}.
\newblock \showarticletitle{Fairness and Sustainability in Multistakeholder Tourism Recommender Systems}. In \bibinfo{booktitle}{\emph{Proceedings of the 31st ACM Conference on User Modeling, Adaptation and Personalization}}. \bibinfo{publisher}{ACM}, \bibinfo{address}{Limassol Cyprus}, \bibinfo{pages}{274–279}.
\newblock
\showISBNx{9781450399326}
\href{https://doi.org/10.1145/3565472.3595607}{doi:\nolinkurl{10.1145/3565472.3595607}}


\bibitem[Bauer et~al\mbox{.}(2024)]%
        {Bauer_Bagchi_Hundogan_Van_Es_2024}
\bibfield{author}{\bibinfo{person}{Christine Bauer}, \bibinfo{person}{Chandni Bagchi}, \bibinfo{person}{Olusanmi~A. Hundogan}, {and} \bibinfo{person}{Karin Van~Es}.} \bibinfo{year}{2024}\natexlab{}.
\newblock \showarticletitle{Where Are the Values? A Systematic Literature Review on News Recommender Systems}.
\newblock \bibinfo{journal}{\emph{ACM Transactions on Recommender Systems}} \bibinfo{volume}{2}, \bibinfo{number}{3} (\bibinfo{date}{Sept.} \bibinfo{year}{2024}), \bibinfo{pages}{1–40}.
\newblock
\href{https://doi.org/10.1145/3654805}{doi:\nolinkurl{10.1145/3654805}}


\bibitem[Burke et~al\mbox{.}(2025)]%
        {Burke_Adomavicius_Bogers_Noia_Kowald_Neidhardt_Özgöbek_Pera_Tintarev_Ziegler_2025}
\bibfield{author}{\bibinfo{person}{Robin Burke}, \bibinfo{person}{Gediminas Adomavicius}, \bibinfo{person}{Toine Bogers}, \bibinfo{person}{Tommaso~Di Noia}, \bibinfo{person}{Dominik Kowald}, \bibinfo{person}{Julia Neidhardt}, \bibinfo{person}{Özlem Özgöbek}, \bibinfo{person}{Maria~Soledad Pera}, \bibinfo{person}{Nava Tintarev}, {and} \bibinfo{person}{Jürgen Ziegler}.} \bibinfo{year}{2025}\natexlab{}.
\newblock \showarticletitle{De-centering the (Traditional) User: Multistakeholder Evaluation of Recommender Systems}.
\newblock  \bibinfo{number}{arXiv:2501.05170} (\bibinfo{date}{Jan.} \bibinfo{year}{2025}).
\newblock
\href{https://doi.org/10.48550/arXiv.2501.05170}{doi:\nolinkurl{10.48550/arXiv.2501.05170}}
\newblock
\shownote{arXiv:2501.05170}.


\bibitem[Burke et~al\mbox{.}(2024)]%
        {dagstuhl_multi_2024}
\bibfield{author}{\bibinfo{person}{Robin Burke}, \bibinfo{person}{Gediminas Adomavicius}, \bibinfo{person}{Toine Bogers}, \bibinfo{person}{Tommaso~Di Noia}, \bibinfo{person}{Dominik Kowald}, \bibinfo{person}{Julia Neidhardt}, \bibinfo{person}{Özlem Özgöbek}, \bibinfo{person}{Maria~Soledad Pera}, {and} \bibinfo{person}{Jürgen Ziegler}.} \bibinfo{year}{2024}\natexlab{}.
\newblock \showarticletitle{Dagstuhl Seminar on Evaluation Perspectives of Recommender Systems: {M}ultistakeholder and Multimethod Evaluation}.
\newblock \bibinfo{journal}{\emph{Dagstuhl Report on Evaluation Perspectives of Recommender Systems: Driving Research and Education}} (\bibinfo{year}{2024}).
\newblock
\urldef\tempurl%
\url{https://doi.org/10.4230/DagRep.14.5.58}
\showURL{%
\tempurl}


\bibitem[Champiri et~al\mbox{.}(2015)]%
        {Champiri_Shahamiri_Salim_2015}
\bibfield{author}{\bibinfo{person}{Zohreh~Dehghani Champiri}, \bibinfo{person}{Seyed~Reza Shahamiri}, {and} \bibinfo{person}{Siti Salwah~Binti Salim}.} \bibinfo{year}{2015}\natexlab{}.
\newblock \showarticletitle{A systematic review of scholar context-aware recommender systems}.
\newblock \bibinfo{journal}{\emph{Expert Systems with Applications}} \bibinfo{volume}{42}, \bibinfo{number}{3} (\bibinfo{year}{2015}), \bibinfo{pages}{1743–1758}.
\newblock
\href{https://doi.org/10.1016/j.eswa.2014.09.017}{doi:\nolinkurl{10.1016/j.eswa.2014.09.017}}


\bibitem[De~Biasio et~al\mbox{.}(2023)]%
        {De_Biasio_Montagna_Aiolli_Navarin_2023}
\bibfield{author}{\bibinfo{person}{Alvise De~Biasio}, \bibinfo{person}{Andrea Montagna}, \bibinfo{person}{Fabio Aiolli}, {and} \bibinfo{person}{Nicolò Navarin}.} \bibinfo{year}{2023}\natexlab{}.
\newblock \showarticletitle{A systematic review of value-aware recommender systems}.
\newblock \bibinfo{journal}{\emph{Expert Systems with Applications}}  \bibinfo{volume}{226} (\bibinfo{year}{2023}), \bibinfo{pages}{120131}.
\newblock
\href{https://doi.org/10.1016/j.eswa.2023.120131}{doi:\nolinkurl{10.1016/j.eswa.2023.120131}}


\bibitem[Deldjoo et~al\mbox{.}(2024)]%
        {Deldjoo_Jannach_Bellogin_Difonzo_Zanzonelli_2024}
\bibfield{author}{\bibinfo{person}{Yashar Deldjoo}, \bibinfo{person}{Dietmar Jannach}, \bibinfo{person}{Alejandro Bellogin}, \bibinfo{person}{Alessandro Difonzo}, {and} \bibinfo{person}{Dario Zanzonelli}.} \bibinfo{year}{2024}\natexlab{}.
\newblock \showarticletitle{Fairness in recommender systems: research landscape and future directions}.
\newblock \bibinfo{journal}{\emph{User Modeling and User-Adapted Interaction}} \bibinfo{volume}{34}, \bibinfo{number}{1} (\bibinfo{year}{2024}), \bibinfo{pages}{59–108}.
\newblock
\href{https://doi.org/10.1007/s11257-023-09364-z}{doi:\nolinkurl{10.1007/s11257-023-09364-z}}


\bibitem[Dobreva et~al\mbox{.}(2012)]%
        {Dobreva_ODwyer_Feliciati_2012}
\bibfield{author}{\bibinfo{person}{Milena Dobreva}, \bibinfo{person}{Andy O’Dwyer}, {and} \bibinfo{person}{Pierluigi Feliciati}.} \bibinfo{year}{2012}\natexlab{}.
\newblock \bibinfo{booktitle}{\emph{Introduction: user studies for digital library development}}.
\newblock \bibinfo{publisher}{Facet}, \bibinfo{pages}{1–18}.
\newblock


\bibitem[Forster et~al\mbox{.}(2025)]%
        {forster2025exploring}
\bibfield{author}{\bibinfo{person}{Andrea Forster}, \bibinfo{person}{Simone Kopeinik}, \bibinfo{person}{Denic Helic}, \bibinfo{person}{Stefan Thalmann}, {and} \bibinfo{person}{Dominik Kowald}.} \bibinfo{year}{2025}\natexlab{}.
\newblock \showarticletitle{Exploring the Effect of Context-Awareness and Popularity Calibration on Popularity Bias in POI Recommendations}.
\newblock \bibinfo{journal}{\emph{arXiv preprint arXiv:2507.03503}} (\bibinfo{year}{2025}).
\newblock
\urldef\tempurl%
\url{https://doi.org/10.48550/arXiv.2507.03503}
\showURL{%
\tempurl}


\bibitem[Färber et~al\mbox{.}(2023)]%
        {Färber_Coutinho_Yuan_2023}
\bibfield{author}{\bibinfo{person}{Michael Färber}, \bibinfo{person}{Melissa Coutinho}, {and} \bibinfo{person}{Shuzhou Yuan}.} \bibinfo{year}{2023}\natexlab{}.
\newblock \showarticletitle{Biases in scholarly recommender systems: impact, prevalence, and mitigation}.
\newblock \bibinfo{journal}{\emph{Scientometrics}} \bibinfo{volume}{128}, \bibinfo{number}{5} (\bibinfo{year}{2023}), \bibinfo{pages}{2703–2736}.
\newblock
\href{https://doi.org/10.1007/s11192-023-04636-2}{doi:\nolinkurl{10.1007/s11192-023-04636-2}}


\bibitem[Gao et~al\mbox{.}(2023)]%
        {Gao_Zheng_Li_Li_Qin_Piao_Quan_Chang_Jin_He_et_al._2023}
\bibfield{author}{\bibinfo{person}{Chen Gao}, \bibinfo{person}{Yu Zheng}, \bibinfo{person}{Nian Li}, \bibinfo{person}{Yinfeng Li}, \bibinfo{person}{Yingrong Qin}, \bibinfo{person}{Jinghua Piao}, \bibinfo{person}{Yuhan Quan}, \bibinfo{person}{Jianxin Chang}, \bibinfo{person}{Depeng Jin}, \bibinfo{person}{Xiangnan He}, {and} \bibinfo{person}{Yong Li}.} \bibinfo{year}{2023}\natexlab{}.
\newblock \showarticletitle{A Survey of Graph Neural Networks for Recommender Systems: Challenges, Methods, and Directions}.
\newblock \bibinfo{journal}{\emph{ACM Transactions on Recommender Systems}} \bibinfo{volume}{1}, \bibinfo{number}{1} (\bibinfo{date}{March} \bibinfo{year}{2023}), \bibinfo{pages}{1–51}.
\newblock
\href{https://doi.org/10.1145/3568022}{doi:\nolinkurl{10.1145/3568022}}


\bibitem[Haberl et~al\mbox{.}(2024)]%
        {Haberl_Fleiß_Kowald_Thalmann_2024}
\bibfield{author}{\bibinfo{person}{Armin Haberl}, \bibinfo{person}{Jürgen Fleiß}, \bibinfo{person}{Dominik Kowald}, {and} \bibinfo{person}{Stefan Thalmann}.} \bibinfo{year}{2024}\natexlab{}.
\newblock \showarticletitle{Take the aTrain. Introducing an interface for the Accessible Transcription of Interviews}.
\newblock \bibinfo{journal}{\emph{Journal of Behavioral and Experimental Finance}}  \bibinfo{volume}{41} (\bibinfo{year}{2024}), \bibinfo{pages}{100891}.
\newblock
\href{https://doi.org/10.1016/j.jbef.2024.100891}{doi:\nolinkurl{10.1016/j.jbef.2024.100891}}


\bibitem[Jannach and Abdollahpouri(2023)]%
        {jannach2023survey}
\bibfield{author}{\bibinfo{person}{Dietmar Jannach} {and} \bibinfo{person}{Himan Abdollahpouri}.} \bibinfo{year}{2023}\natexlab{}.
\newblock \showarticletitle{A survey on multi-objective recommender systems}.
\newblock \bibinfo{journal}{\emph{Frontiers in big Data}}  \bibinfo{volume}{6} (\bibinfo{year}{2023}), \bibinfo{pages}{1157899}.
\newblock
\urldef\tempurl%
\url{https://doi.org/10.3389/fdata.2023.1157899}
\showURL{%
\tempurl}


\bibitem[Jannach and Zanker(2024)]%
        {Jannach_Zanker_2024}
\bibfield{author}{\bibinfo{person}{Dietmar Jannach} {and} \bibinfo{person}{Markus Zanker}.} \bibinfo{year}{2024}\natexlab{}.
\newblock \showarticletitle{A Survey on Intent-aware Recommender Systems}.
\newblock  \bibinfo{number}{arXiv:2406.16350} (\bibinfo{date}{Oct.} \bibinfo{year}{2024}).
\newblock
\href{https://doi.org/10.48550/arXiv.2406.16350}{doi:\nolinkurl{10.48550/arXiv.2406.16350}}
\newblock
\shownote{arXiv:2406.16350}.


\bibitem[Klimashevskaia et~al\mbox{.}(2024)]%
        {Klimashevskaia_Jannach_Elahi_Trattner_2024}
\bibfield{author}{\bibinfo{person}{Anastasiia Klimashevskaia}, \bibinfo{person}{Dietmar Jannach}, \bibinfo{person}{Mehdi Elahi}, {and} \bibinfo{person}{Christoph Trattner}.} \bibinfo{year}{2024}\natexlab{}.
\newblock \showarticletitle{A survey on popularity bias in recommender systems}.
\newblock \bibinfo{journal}{\emph{User Modeling and User-Adapted Interaction}} \bibinfo{volume}{34}, \bibinfo{number}{5} (\bibinfo{year}{2024}), \bibinfo{pages}{1777–1834}.
\newblock
\href{https://doi.org/10.1007/s11257-024-09406-0}{doi:\nolinkurl{10.1007/s11257-024-09406-0}}


\bibitem[Lacic et~al\mbox{.}(2015)]%
        {lacic2015tackling}
\bibfield{author}{\bibinfo{person}{Emanuel Lacic}, \bibinfo{person}{Dominik Kowald}, \bibinfo{person}{Matthias Traub}, \bibinfo{person}{Granit Luzhnica}, \bibinfo{person}{J{\"o}rg~Peter Simon}, {and} \bibinfo{person}{Elisabeth Lex}.} \bibinfo{year}{2015}\natexlab{}.
\newblock \showarticletitle{Tackling Cold-Start Users in Recommender Systems with Indoor Positioning Systems}. In \bibinfo{booktitle}{\emph{9th ACM Conference on Recommender Systems}}. ACM.
\newblock
\urldef\tempurl%
\url{https://ceur-ws.org/Vol-1441/recsys2015_poster21.pdf}
\showURL{%
\tempurl}


\bibitem[Li et~al\mbox{.}(2024)]%
        {Li_Zhang_Wang_2024}
\bibfield{author}{\bibinfo{person}{Wenqi Li}, \bibinfo{person}{Pengyi Zhang}, {and} \bibinfo{person}{Jun Wang}.} \bibinfo{year}{2024}\natexlab{}.
\newblock \showarticletitle{Analysing humanities scholars’ data seeking behaviour patterns using Ellis’ model}.
\newblock \bibinfo{journal}{\emph{Information Research an international electronic journal}} \bibinfo{volume}{29}, \bibinfo{number}{2} (\bibinfo{date}{June} \bibinfo{year}{2024}), \bibinfo{pages}{401–418}.
\newblock
\href{https://doi.org/10.47989/ir292835}{doi:\nolinkurl{10.47989/ir292835}}


\bibitem[Lian et~al\mbox{.}(2021)]%
        {Lian_Batal_Liu_Soni_Kang_Wang_Xie_2021}
\bibfield{author}{\bibinfo{person}{Jianxun Lian}, \bibinfo{person}{Iyad Batal}, \bibinfo{person}{Zheng Liu}, \bibinfo{person}{Akshay Soni}, \bibinfo{person}{Eun~Yong Kang}, \bibinfo{person}{Yajun Wang}, {and} \bibinfo{person}{Xing Xie}.} \bibinfo{year}{2021}\natexlab{}.
\newblock \showarticletitle{Multi-Interest-Aware User Modeling for Large-Scale Sequential Recommendations}.
\newblock  \bibinfo{number}{arXiv:2102.09211} (\bibinfo{date}{May} \bibinfo{year}{2021}).
\newblock
\href{https://doi.org/10.48550/arXiv.2102.09211}{doi:\nolinkurl{10.48550/arXiv.2102.09211}}
\newblock
\shownote{arXiv:2102.09211}.


\bibitem[Marchionini(2024)]%
        {Marchionini_2024}
\bibfield{author}{\bibinfo{person}{Gary Marchionini}.} \bibinfo{year}{2024}\natexlab{}.
\newblock \showarticletitle{Information and library professionals’ roles and responsibilities in an AI ‐augmented world}.
\newblock \bibinfo{journal}{\emph{Journal of the Association for Information Science and Technology}} \bibinfo{volume}{75}, \bibinfo{number}{8} (\bibinfo{year}{2024}), \bibinfo{pages}{865–868}.
\newblock
\href{https://doi.org/10.1002/asi.24930}{doi:\nolinkurl{10.1002/asi.24930}}


\bibitem[Marchionini et~al\mbox{.}(2003)]%
        {Marchionini_Plaisant_Komlodi_2003}
\bibfield{author}{\bibinfo{person}{Gary Marchionini}, \bibinfo{person}{Catherine Plaisant}, {and} \bibinfo{person}{Anita Komlodi}.} \bibinfo{year}{2003}\natexlab{}.
\newblock \bibinfo{booktitle}{\emph{The People in Digital Libraries: Multifaceted Approaches to Assessing Needs and Impact}}.
\newblock \bibinfo{publisher}{The MIT Press}, \bibinfo{pages}{119–160}.
\newblock
\showISBNx{9780262255745}
\href{https://doi.org/10.7551/mitpress/2424.003.0009}{doi:\nolinkurl{10.7551/mitpress/2424.003.0009}}


\bibitem[Radford et~al\mbox{.}(2022)]%
        {Radford_Kim_Xu_Brockman_McLeavey_Sutskever_2022}
\bibfield{author}{\bibinfo{person}{Alec Radford}, \bibinfo{person}{Jong~Wook Kim}, \bibinfo{person}{Tao Xu}, \bibinfo{person}{Greg Brockman}, \bibinfo{person}{Christine McLeavey}, {and} \bibinfo{person}{Ilya Sutskever}.} \bibinfo{year}{2022}\natexlab{}.
\newblock \showarticletitle{Robust Speech Recognition via Large-Scale Weak Supervision}.
\newblock  \bibinfo{number}{arXiv:2212.04356} (\bibinfo{date}{Dec.} \bibinfo{year}{2022}).
\newblock
\href{https://doi.org/10.48550/arXiv.2212.04356}{doi:\nolinkurl{10.48550/arXiv.2212.04356}}
\newblock
\shownote{arXiv:2212.04356}.


\bibitem[Rezk et~al\mbox{.}(2024)]%
        {Rezk_Simkute_Luger_Vines_Elsden_Evans_Jones_2024}
\bibfield{author}{\bibinfo{person}{Anna~Marie Rezk}, \bibinfo{person}{Auste Simkute}, \bibinfo{person}{Ewa Luger}, \bibinfo{person}{John Vines}, \bibinfo{person}{Chris Elsden}, \bibinfo{person}{Michael Evans}, {and} \bibinfo{person}{Rhianne Jones}.} \bibinfo{year}{2024}\natexlab{}.
\newblock \showarticletitle{Agency Aspirations: Understanding Users’ Preferences And Perceptions Of Their Role In Personalised News Curation}. In \bibinfo{booktitle}{\emph{Proceedings of the CHI Conference on Human Factors in Computing Systems}}. \bibinfo{publisher}{ACM}, \bibinfo{address}{Honolulu HI USA}, \bibinfo{pages}{1–16}.
\newblock
\showISBNx{9798400703300}
\href{https://doi.org/10.1145/3613904.3642634}{doi:\nolinkurl{10.1145/3613904.3642634}}


\bibitem[Semmelrock et~al\mbox{.}(2025)]%
        {semmelrock2025reproducibility}
\bibfield{author}{\bibinfo{person}{Harald Semmelrock}, \bibinfo{person}{Tony Ross-Hellauer}, \bibinfo{person}{Simone Kopeinik}, \bibinfo{person}{Dieter Theiler}, \bibinfo{person}{Armin Haberl}, \bibinfo{person}{Stefan Thalmann}, {and} \bibinfo{person}{Dominik Kowald}.} \bibinfo{year}{2025}\natexlab{}.
\newblock \showarticletitle{Reproducibility in machine-learning-based research: Overview, barriers, and drivers}.
\newblock \bibinfo{journal}{\emph{AI Magazine}} \bibinfo{volume}{46}, \bibinfo{number}{2} (\bibinfo{year}{2025}), \bibinfo{pages}{e70002}.
\newblock
\urldef\tempurl%
\url{https://doi.org/10.1002/aaai.70002}
\showURL{%
\tempurl}


\bibitem[Shehzad et~al\mbox{.}(2025)]%
        {Shehzad_Dacrema_Jannach_2025}
\bibfield{author}{\bibinfo{person}{Faisal Shehzad}, \bibinfo{person}{Maurizio~Ferrari Dacrema}, {and} \bibinfo{person}{Dietmar Jannach}.} \bibinfo{year}{2025}\natexlab{}.
\newblock \showarticletitle{A Worrying Reproducibility Study of Intent-Aware Recommendation Models}.
\newblock  \bibinfo{number}{arXiv:2501.10143} (\bibinfo{date}{Jan.} \bibinfo{year}{2025}).
\newblock
\href{https://doi.org/10.48550/arXiv.2501.10143}{doi:\nolinkurl{10.48550/arXiv.2501.10143}}
\newblock
\shownote{arXiv:2501.10143}.


\bibitem[Smith et~al\mbox{.}(2024)]%
        {Smith_Satwani_Burke_Fiesler_2024}
\bibfield{author}{\bibinfo{person}{Jessie~J. Smith}, \bibinfo{person}{Aishwarya Satwani}, \bibinfo{person}{Robin Burke}, {and} \bibinfo{person}{Casey Fiesler}.} \bibinfo{year}{2024}\natexlab{}.
\newblock \showarticletitle{Recommend Me? Designing Fairness Metrics with Providers}. In \bibinfo{booktitle}{\emph{The 2024 ACM Conference on Fairness, Accountability, and Transparency}}. \bibinfo{publisher}{ACM}, \bibinfo{address}{Rio de Janeiro Brazil}, \bibinfo{pages}{2389–2399}.
\newblock
\showISBNx{9798400704505}
\href{https://doi.org/10.1145/3630106.3659044}{doi:\nolinkurl{10.1145/3630106.3659044}}


\bibitem[Starke et~al\mbox{.}(2024)]%
        {Starke_Vrijenhoek_Michiels_Kruse_Tintarev_2024}
\bibfield{author}{\bibinfo{person}{Alain~Dominique Starke}, \bibinfo{person}{Sanne Vrijenhoek}, \bibinfo{person}{Lien Michiels}, \bibinfo{person}{Johannes Kruse}, {and} \bibinfo{person}{Nava Tintarev}.} \bibinfo{year}{2024}\natexlab{}.
\newblock \bibinfo{booktitle}{\emph{Report on NORMalize: The Second Workshop on the Normative Design and Evaluation of Recommender Systems}}.
\newblock
\showISBNx{9780609780190}
\urldef\tempurl%
\url{https://bora.uib.no/bora-xmlui/handle/11250/3185749}
\showURL{%
\tempurl}


\bibitem[Stray et~al\mbox{.}(2024)]%
        {Stray_Halevy_Assar_Hadfield-Menell_Boutilier_Ashar_Bakalar_Beattie_Ekstrand_Leibowicz_et_al._2024}
\bibfield{author}{\bibinfo{person}{Jonathan Stray}, \bibinfo{person}{Alon Halevy}, \bibinfo{person}{Parisa Assar}, \bibinfo{person}{Dylan Hadfield-Menell}, \bibinfo{person}{Craig Boutilier}, \bibinfo{person}{Amar Ashar}, \bibinfo{person}{Chloe Bakalar}, \bibinfo{person}{Lex Beattie}, \bibinfo{person}{Michael Ekstrand}, \bibinfo{person}{Claire Leibowicz}, \bibinfo{person}{Connie Moon~Sehat}, \bibinfo{person}{Sara Johansen}, \bibinfo{person}{Lianne Kerlin}, \bibinfo{person}{David Vickrey}, \bibinfo{person}{Spandana Singh}, \bibinfo{person}{Sanne Vrijenhoek}, \bibinfo{person}{Amy Zhang}, \bibinfo{person}{McKane Andrus}, \bibinfo{person}{Natali Helberger}, \bibinfo{person}{Polina Proutskova}, \bibinfo{person}{Tanushree Mitra}, {and} \bibinfo{person}{Nina Vasan}.} \bibinfo{year}{2024}\natexlab{}.
\newblock \showarticletitle{Building Human Values into Recommender Systems: An Interdisciplinary Synthesis}.
\newblock \bibinfo{journal}{\emph{ACM Transactions on Recommender Systems}} \bibinfo{volume}{2}, \bibinfo{number}{3} (\bibinfo{date}{Sept.} \bibinfo{year}{2024}), \bibinfo{pages}{1–57}.
\newblock
\href{https://doi.org/10.1145/3632297}{doi:\nolinkurl{10.1145/3632297}}


\bibitem[Van Den~Bogaert et~al\mbox{.}(2024)]%
        {Van_Den_Bogaert_Geerts_Harambam_2024}
\bibfield{author}{\bibinfo{person}{Lawrence Van Den~Bogaert}, \bibinfo{person}{David Geerts}, {and} \bibinfo{person}{Jaron Harambam}.} \bibinfo{year}{2024}\natexlab{}.
\newblock \showarticletitle{Putting a Human Face on the Algorithm: Co-Designing Recommender Personae to Democratize News Recommender Systems}.
\newblock \bibinfo{journal}{\emph{Digital Journalism}} \bibinfo{volume}{12}, \bibinfo{number}{8} (\bibinfo{date}{Sept.} \bibinfo{year}{2024}), \bibinfo{pages}{1097–1117}.
\newblock
\href{https://doi.org/10.1080/21670811.2022.2097101}{doi:\nolinkurl{10.1080/21670811.2022.2097101}}


\bibitem[Vandenbroucke and Smets(2024)]%
        {Vandenbroucke_Smets_2024}
\bibfield{author}{\bibinfo{person}{Hanne Vandenbroucke} {and} \bibinfo{person}{Annelien Smets}.} \bibinfo{year}{2024}\natexlab{}.
\newblock \showarticletitle{It’s (not) all about that CTR: A Multi-Stakeholder Perspective on News Recommender Metrics}. In \bibinfo{booktitle}{\emph{18th ACM Conference on Recommender Systems}}. \bibinfo{publisher}{ACM}, \bibinfo{address}{Bari Italy}, \bibinfo{pages}{999–1003}.
\newblock
\showISBNx{9798400705052}
\href{https://doi.org/10.1145/3640457.3688183}{doi:\nolinkurl{10.1145/3640457.3688183}}


\bibitem[Wang et~al\mbox{.}(2022)]%
        {Wang_Zhang_Hu_Zhang_Wang_Aggarwal_2022}
\bibfield{author}{\bibinfo{person}{Shoujin Wang}, \bibinfo{person}{Qi Zhang}, \bibinfo{person}{Liang Hu}, \bibinfo{person}{Xiuzhen Zhang}, \bibinfo{person}{Yan Wang}, {and} \bibinfo{person}{Charu Aggarwal}.} \bibinfo{year}{2022}\natexlab{}.
\newblock \showarticletitle{Sequential/Session-based Recommendations: Challenges, Approaches, Applications and Opportunities}. In \bibinfo{booktitle}{\emph{Proceedings of the 45th International ACM SIGIR Conference on Research and Development in Information Retrieval}}. \bibinfo{publisher}{ACM}, \bibinfo{address}{Madrid Spain}, \bibinfo{pages}{3425–3428}.
\newblock
\showISBNx{9781450387323}
\href{https://doi.org/10.1145/3477495.3532685}{doi:\nolinkurl{10.1145/3477495.3532685}}


\bibitem[Wardatzky et~al\mbox{.}(2025)]%
        {Wardatzky_Inel_Rossetto_Bernstein_2025}
\bibfield{author}{\bibinfo{person}{Kathrin Wardatzky}, \bibinfo{person}{Oana Inel}, \bibinfo{person}{Luca Rossetto}, {and} \bibinfo{person}{Abraham Bernstein}.} \bibinfo{year}{2025}\natexlab{}.
\newblock \showarticletitle{Whom do Explanations Serve? A Systematic Literature Survey of User Characteristics in Explainable Recommender Systems Evaluation}.
\newblock \bibinfo{journal}{\emph{ACM Transactions on Recommender Systems}} (\bibinfo{date}{Feb.} \bibinfo{year}{2025}), \bibinfo{pages}{3716394}.
\newblock
\href{https://doi.org/10.1145/3716394}{doi:\nolinkurl{10.1145/3716394}}


\bibitem[Wecker et~al\mbox{.}(2023)]%
        {Wecker_Kuflik_Goldberg_Lanir_Tabashi_2023}
\bibfield{author}{\bibinfo{person}{Alan~Jay Wecker}, \bibinfo{person}{Tsvi Kuflik}, \bibinfo{person}{Tsafrir Goldberg}, \bibinfo{person}{Joel Lanir}, {and} \bibinfo{person}{Tal Tabashi}.} \bibinfo{year}{2023}\natexlab{}.
\newblock \showarticletitle{Using Recommendations to Affect Social Change in Cultural Heritage: Should We and How?}. In \bibinfo{booktitle}{\emph{Adjunct Proceedings of the 31st ACM Conference on User Modeling, Adaptation and Personalization}}. \bibinfo{publisher}{ACM}, \bibinfo{address}{Limassol Cyprus}, \bibinfo{pages}{419–421}.
\newblock
\showISBNx{9781450398916}
\href{https://doi.org/10.1145/3563359.3596663}{doi:\nolinkurl{10.1145/3563359.3596663}}


\bibitem[Zangerle and Bauer(2023)]%
        {Zangerle_Bauer_2023}
\bibfield{author}{\bibinfo{person}{Eva Zangerle} {and} \bibinfo{person}{Christine Bauer}.} \bibinfo{year}{2023}\natexlab{}.
\newblock \showarticletitle{Evaluating Recommender Systems: Survey and Framework}.
\newblock \bibinfo{journal}{\emph{Comput. Surveys}} \bibinfo{volume}{55}, \bibinfo{number}{8} (\bibinfo{date}{Aug.} \bibinfo{year}{2023}), \bibinfo{pages}{1–38}.
\newblock
\href{https://doi.org/10.1145/3556536}{doi:\nolinkurl{10.1145/3556536}}


\bibitem[Zhang et~al\mbox{.}(2023)]%
        {Zhang_Patra_Yaseen_Zhu_Sabharwal_Roberts_Cao_Wu_2023}
\bibfield{author}{\bibinfo{person}{Zitong Zhang}, \bibinfo{person}{Braja~Gopal Patra}, \bibinfo{person}{Ashraf Yaseen}, \bibinfo{person}{Jie Zhu}, \bibinfo{person}{Rachit Sabharwal}, \bibinfo{person}{Kirk Roberts}, \bibinfo{person}{Tru Cao}, {and} \bibinfo{person}{Hulin Wu}.} \bibinfo{year}{2023}\natexlab{}.
\newblock \showarticletitle{Scholarly recommendation systems: a literature survey}.
\newblock \bibinfo{journal}{\emph{Knowledge and Information Systems}} \bibinfo{volume}{65}, \bibinfo{number}{11} (\bibinfo{year}{2023}), \bibinfo{pages}{4433–4478}.
\newblock
\href{https://doi.org/10.1007/s10115-023-01901-x}{doi:\nolinkurl{10.1007/s10115-023-01901-x}}


\bibitem[Zhao et~al\mbox{.}(2025)]%
        {Zhao_Wang_Chen_Gao_Wang_Li_Jia_Liu_Guo_Tang_2025}
\bibfield{author}{\bibinfo{person}{Xiangyu Zhao}, \bibinfo{person}{Yichao Wang}, \bibinfo{person}{Bo Chen}, \bibinfo{person}{Jingtong Gao}, \bibinfo{person}{Yuhao Wang}, \bibinfo{person}{Xiaopeng Li}, \bibinfo{person}{Pengyue Jia}, \bibinfo{person}{Qidong Liu}, \bibinfo{person}{Huifeng Guo}, {and} \bibinfo{person}{Ruiming Tang}.} \bibinfo{year}{2025}\natexlab{}.
\newblock \showarticletitle{Joint Modeling in Recommendations: A Survey}.
\newblock  \bibinfo{number}{arXiv:2502.21195} (\bibinfo{date}{Feb.} \bibinfo{year}{2025}).
\newblock
\href{https://doi.org/10.48550/arXiv.2502.21195}{doi:\nolinkurl{10.48550/arXiv.2502.21195}}
\newblock
\shownote{arXiv:2502.21195}.


\end{thebibliography}

\end{document}